\documentclass[twocolumn,showpacs,preprintnumbers,amsmath,amssymb]{revtex4}
\usepackage{graphicx}
\usepackage{dcolumn}
\usepackage{color}
\usepackage{bm}
\begin{document}
\title {Diffusion Enhancement in a Periodic Potential under High-Frequency Space-Dependent Forcing}
\author{ Malay Bandyopadhyay$^{1}$, Sushanta Dattagupta$^{1}$ and Monamie Sanyal$^{2}$}
\affiliation{$^{1}$S.N. Bose National Centre for Basic
Sciences,JD Block, Sector III, Salt Lake, Kolkata 700098, India.\\
$^{2}$Department of Physics, Indian Institute of Technology, Kanpur.}
\date{\today}
\begin{abstract}
We study the long-time behavior of underdamped Brownian particle moving through a viscous medium and in a systematic potential, when it is subjected to a space-dependent high-frequency periodic force. When the frequency is very large, much larger than all other relevant system-frequencies, there is a Kapitsa time-window wherein the effect of frequency dependent forcing can be replaced by a static effective potential. Our new analysis includes the case when the forcing, in addition to being frequency-dependent, is space-dependent as well. The results of the Kapitsa analysis then lead to additional contributions to the effective potential. These are applied to the numerical calculation of the diffusion coefficient (D) for a Brownian particle moving in a periodic potential. Presented are numerical results, which are in excellent agreement with theoretical predictions and which indicate a significant enhancement of D due to the space-dependent forcing terms. In addition we study the transport property (current) of underdamped Brownian particles in a ratchet potential. 
\end{abstract}
\maketitle
{\section {Introduction}}
Brownian motion in periodic structures has various applications to condensed matter physics, nanotechnology and molecular biology \cite{reim1,land,brau}. Adding noise to deterministic nonlinear dynamics has led to interesting and important phenomena such as stochastic resonance \cite{hanggi1}, Brownian motors and chaotic ratchet transport \cite{hanggi2}, resonant activation \cite{doer}, noise induced phase transition \cite{vanden1,vanden2}, etc.. Thermal diffusion of a Brownian particle which we will discuss here is of great interest in numerous other contexts, namely Josephson junction \cite{baron}, rotating dipoles in external fields \cite{regue}, superionic conductors \cite{fulde}, synchronization phenomena \cite{lind}, diffusion on crystal surfaces \cite{fren}, particle separation by electrophoresis \cite{ajdari}, and biophysical processes such as intracellular transport \cite{reim2}.\\
\noindent
In this paper we focus on the underdamped motion of a Brownian particle which feels viscous forces and random impulses from the surrounding medium and is confined by a potential well. Our primary interest is to study the effect of an externally applied position dependent driving force that is periodic in time. The frequency is much larger than all other relevant frequencies of the system. Hence we can apply the usual Kapitsa analysis for high frequency oscillating fields \cite{kapitsa}. It has earlier been shown that on time scales larger than the period of perturbation the dynamics is equivalent to one in which the periodic perturbation can be replaced by a time independent effective potential \cite{dutta,sarkar}. Ref. \cite{sarkar} treats the overdamped Brownian motion, whereas we deal with underdamped motion of the Brownian particle. Dutta {\em {et al.}} \cite{dutta} have based their analysis on the Fokker-Planck equation approach. Here we provide an alternative derivation of the main results through the Langevin dynamics, which is more straightforward.\\
\noindent
Further we extend  Kapitsa's analysis by using the additional contributions to the effective potential arising from the space dependent periodic force for the calculation of the thermal diffusion coefficient \cite{reim3,mach}. The central result of our present paper is that the effective diffusion coefficient of an underdamped Brownian particle in a periodic potential in presence of an externally applied space-dependent periodic force can become arbitrarily larger than in the presence of thermal noise alone. Furthermore, certain features of the ratchet mechanism \cite{hanggi3} are relevant in this context in terms of transport properties(currents).\\
\noindent
With the preceding background the paper is organized as follows. In Sec.II we introduce the model, the numerical scheme and the basic quantities of interest, namely, the effective potential, the effective diffusion coefficient, and the average particle current. In Sec.III we develop the necessary formalism to address rapidly periodic drive and arrive at the perturbative effective potential. In Sec.IV we discuss the numerical results on the effective diffusion coefficient and average particle current. The summary remarks, discussion and conclusion of our findings are presented in Sec.V.\\
{\section {Model}}
The stated Brownian dynamics is governed by the Langevin equation
\begin{equation}
m\ddot{x} = -\gamma\dot{x}-\frac{\partial}{\partial x}U(x) + F(x,t) + \eta(t),
\end{equation}
where m is the mass of the Brownian particle, $\gamma$ is the friction coefficient, U(x) is the confining potential, F(x,t) is the periodic driving force with a period $\tau$ and F(x,t) = F(x,t+$\tau$). Thermal fluctuations are modeled by the zero mean $\delta$-correlated white noise $\eta(t)$ i.e. $<\eta(t)>$ = 0 and $<\eta(t)\eta(t^{\prime})> = 2\gamma\beta^{-1}\delta(t-t^{\prime})$, where $\beta =(k_BT)^{-1}$, $k_B$ being the Boltzmann constant and T is the temperature. Our goal is to show that, on time scales larger than $\tau$, the dynamics can be mapped onto a modified Langevin dynamics in which the periodic forcing is absent but the potential U(x) can be replaced by a suitable effective potential $U_{eff}(x)$. The methdology we follow is based on Kapitsa's treatment for high frequency oscillating field in parametric oscillations \cite{landau}. We derive the form of the effective potential upto second order in $\xi$ (expansion parameter which is basically inverse of the square of the oscillating frequency) in Sec.III.\\
\noindent
We also study the transport properties and diffusion coefficient for ratchet like systems. The corresponding Langevin dynamics is governed by the equation
\begin{equation}
m\ddot{x} + \gamma\dot{x} = -V^{\prime}(x) + A(x)\cos(\Omega t) +\sqrt{2\gamma k_BT}\eta(t),
\end{equation}
where V(x) is the periodic potential with period L i.e. V(x) = V(x+L) and the prime denotes 1st derivative of V(x) with respect to x. In our case $V(x) = -\sin(x) - \mu\sin(2x)$ and $\mu = \frac{1}{4}$ throughout this work.      
\begin{figure}[h]
{\rotatebox{270}{\resizebox{5cm}{6cm}{\includegraphics{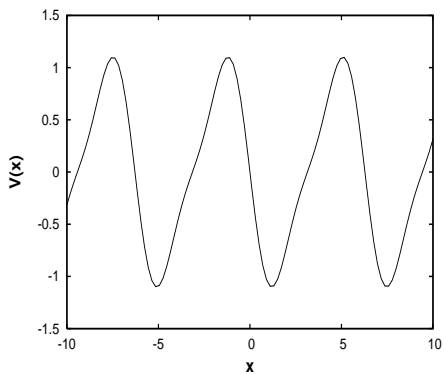}}}}
\caption{The periodic potential V(x).}
\end{figure}
We define the above dynamics as the original dynamics. Following Kapitsa's treatment we derive in the sequel an effective potential for which the dynamics is governed by the equation
\begin{equation}
m\ddot{x} + \gamma\dot{x} = -V_{eff}^{\prime}(x) +\sqrt{2\gamma k_BT}\eta(t).
\end{equation}

\begin{figure}[h]
{\rotatebox{0}{\resizebox{6cm}{5cm}{\includegraphics{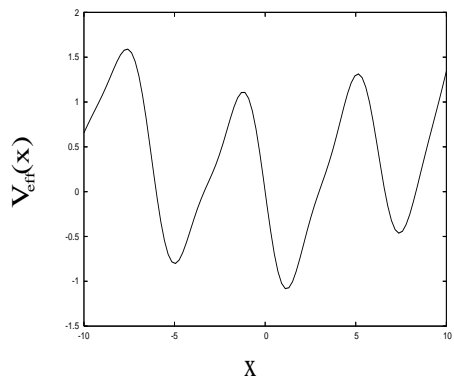}}}}
\caption{The effective potential $V_{eff}(x)$.}
\end{figure}
The first basic quantity of interest in the transport process is the average particle current defined as 
\begin{equation}
<\dot{x}> = \lim_{t \to \infty} \frac{<x(t)>}{t}.
\end{equation}
The other quantity of important interest is the effective diffusion coefficient
\begin{equation}
D = \lim_{t \to \infty} \frac{<x^{2}(t)> - <x(t)>^{2}}{2t}.
\end{equation}
Exact analytical results for D are known for two special cases. First in the absence of the periodic potential we have the famous Einstein's relation $ D = \frac{k_BT}{\gamma}$. Second is the case in which the external field $A(x)\cos(\Omega t)$ is absent. Calculation of the diffusion coefficient in the presence of both externally applied space-dependent periodic force and arbitrary periodic potential is not analytically possible. Hence we solve Eq.(2) and Eq.(3) numerically. To compute x(t) we need to transform Eq.(2) and Eq.(3) into dimensionless Langevin equations as follows:
\begin{eqnarray}
\ddot{x} + b\dot{x}& = & -\hat{V}^{\prime}(x) + a(x)\cos(\omega t) + \sqrt{2bD_0}\hat{\eta}\\
\ddot{x} + b\dot{x}& = & -\hat{V}_{eff}^{\prime}(x) + \sqrt{2bD_0}\hat{\eta},
\end{eqnarray}
where Eq.(6) denotes original dynamics, whereas Eq.(7) is for effective dynamics. In writing the above equations we define the following dimensionless quantities:
\begin{enumerate}
\item the rescaled friction coefficient $b=\frac{\gamma}{m}\tau_0$ which is the ratio of the two characteristic time scales, $\tau_0$ and $\tau_V$. $\tau_0$ is determined from Newton's equation $m\ddot{x} = -V^{\prime}(x)$, by balancing the two forces $\frac{mL}{\tau_0^{2}} = \frac{\Delta V}{L}$ yielding $\tau_0^2 = \frac{mL^2}{\Delta V}$, whereas $\tau_V = \frac{m}{\gamma}$, 
\item the rescaled potential $\hat{V}(x) = \frac{V(x)}{\Delta V}$, assuming period 1 and barrier height $\Delta V = 1$, 
\item the rescaled force strength $ a = \frac{AL}{\Delta V}$ with the frequency $\omega = \Omega\tau_0$, 
\item the rescaled white noise forces $\hat{\eta}(t)$ obeying \\$<\hat{\eta}(t)\hat{\eta}(t^{\prime})> = \delta(t - t^{\prime})$  with rescaled noise intensity $ D_0 = \frac{k_BT}{\Delta V}$.
\end{enumerate}
{\section {Effective Potential}}
As mentioned earlier our focus is on the result that on time scales larger than the period of perturbation, the dynamics is equivalent to one in which the time dependent periodic perturbation can be replaced by a time independent effective potential \cite{jung,landau}. In Sarkar and Dattagupta \cite{sarkar}, it has been shown in great detail that the expression of the effective potential does not alter in the presence of noise. We presume and verify that this result is true even when the forcing term is space dependent. Further, unlike \cite{sarkar} we treat the underdamped case from which all the results of \cite{sarkar} can be obtained as a limit.\\
{\subsection{First Order Correction}}
It is evident from the nature of the field in which the particle moves that it will traverse a smooth path and at the same time will execute small noisy fluctuations about that path. Accordingly, we represent the function x(t) as a sum:
\begin{equation}
x(t) = X(t) + \xi(t),
\end{equation} 
where X(t) is a slow variable and $\xi(X,t)$ is a fast variable. The following transformations then follow:
\begin{eqnarray}
\dot{x} & = & \dot{X} + \dot{\xi}(X,t)
\nonumber \\
\ddot{x} & = & \ddot{X} + \ddot{\xi}(X,t)
\nonumber  \\
\frac{\partial}{\partial x} & = & \frac{1}{1+\xi}\frac{\partial}{\partial X}.
\end{eqnarray}
Now setting the noise term to zero and putting the above transformations in Eq.(1) we obtain
\begin{eqnarray}
m\Big(\ddot{X}(t)+\ddot{\xi}(X,t)\Big)& = &-\gamma\Big(\dot{X}(t)+\dot{\xi}(X,t)\Big)+ F(X+\xi,t) 
\nonumber \\
& &-\frac{1}{1+\xi}\frac{\partial}{\partial X}U(X+\xi).
\end{eqnarray}
To find the effective potential experienced by the particle correct to first order, we perform a Taylor series expansion of Eq.(10) upto first order. Thus,
\begin{eqnarray}
m\ddot{X}(t)+m\ddot{\xi}(X,t)& = &-\gamma\dot{X}(t)-\gamma\dot{\xi}(X,t) 
\nonumber \\ 
&-&\frac{1}{1+\xi}\frac{\partial}{\partial X}(U(X)+\xi U^{\prime}(X))
\nonumber \\
&+&  F(X,t)+\xi F^{\prime}(X,t).
\end{eqnarray}
The above Eq.(11) involves both `fluctuating' and `smooth' terms on the left and right sides which must be separately equal. For the fluctuating terms we can simply put
\begin{equation}
m\ddot{\xi}(t) + \gamma\dot{\xi}(t) = F(X,t),
\end{equation}
where we take \cite{dutta} $F(X,t) = f(X)\cos(\omega t) + g(X)\sin(\omega t)$. Solving Eq.(12) we obtain,
\begin{eqnarray}
\xi(X,t) & = & \frac{1}{m(\omega^2+\frac{\gamma^2}{m^2})}\Big\lbrack\Big(f(X)+\frac{\gamma}{m\omega}g(X)\Big)\cos(\omega t)
\nonumber \\
&+& \Big(g(X)-\frac{\gamma}{m\omega}f(X)\Big)\sin(\omega t)\Big\rbrack.
\end{eqnarray}
\noindent
Since $\omega$ is large, $\frac{1}{1+\xi} \simeq (1-\xi)$ is effective. Next combining Eq.(12) with Eq.(11) and then averaging over a time period we finally obtain
\begin{equation}
m\ddot{X}(t) + \gamma\dot{X}(t) = -\frac{\partial U(X)}{\partial X} + <\xi F^{\prime}(X,t)>,
\end{equation}
where
\begin{eqnarray}
<\xi F^{\prime}(X,t)> & = & -\frac{1}{4m\Big(\omega^2+\frac{\gamma^2}{m^2}\Big)}\Big\lbrack\frac{\partial}{\partial X}(f^2(X) + g^2(X))
\nonumber \\
& & + \frac{2\gamma}{m\omega}(f^{\prime}(X)g(X)-g^{\prime}(X)f(X))\Big\rbrack.
\end{eqnarray}
With the help of Eq.(15) we can rewrite Eq.(14) as follows
\begin{eqnarray}
m\ddot{X}(t) + \gamma\dot{X}(t)& = &-\frac{\partial U(X)}{\partial X}
\nonumber \\
&-& \frac{1}{4m\Big(\omega^2+\frac{\gamma^2}{m^2}\Big)}\Big\lbrack\frac{\partial}{\partial X}(f^2(X) + g^2(X))
\nonumber \\
&+&\frac{2\gamma}{m\omega}(f^{\prime}(X)g(X)-g^{\prime}(X)f(X))\Big\rbrack, 
\end{eqnarray}
or,
\begin{equation}
m\ddot{X}(t) + \gamma\dot{X}(t) = -\frac{\partial U_{eff}(X)}{\partial X},
\end{equation}
with $U_{eff}(X) = U(X) + U^1(X)$, where
\begin{eqnarray}
U^1(X) & = &  \frac{1}{4m\Big(\omega^2+\frac{\gamma^2}{m^2}\Big)}\Big\lbrack\frac{\partial}{\partial X}(f^2(X) + g^2(X))
\nonumber \\
&+&\frac{2\gamma}{m\omega}\int^{X}dy(f^{\prime}(y)g(y)-g^{\prime}(y)f(y))\Big\rbrack.
\end{eqnarray}
It is clear that $U^1(X)$ vanishes for space-independent forcing.
{\subsection{Second Order Correction}}
To find the second order correction term in the effective potential the transformation equations given in Eq.(9) have to be modified as
\begin{eqnarray}
x & = & X + \xi(X,t) + \xi(X,t)\xi^{\prime}(X,t)
\nonumber \\
\dot{x} & = & \dot{X} + \dot{\xi} + \dot{\xi}\xi^{\prime} + \xi\dot{\xi}^{\prime}
\nonumber \\
\ddot{x} & = & \ddot{X} + \ddot{\xi} + \ddot{\xi}\xi^{\prime} + \xi\ddot{\xi}^{\prime}+2\dot{\xi}\dot{\xi}^{\prime}
\nonumber \\
\frac{\partial}{\partial x} & = & \frac{1}{1+\xi^{\prime}+\xi^{\prime 2}+\xi\xi^{\prime\prime}}\frac{\partial}{\partial X}
\end{eqnarray}
Putting the above transformation in Eq.(1) and retaining terms upto seond order in $\xi$ (O$(\xi^2)$) we derive
\begin{eqnarray*}
m\Big(\ddot{X} + \ddot{\xi} + \ddot{\xi}\xi^{\prime} + \xi\ddot{\xi}^{\prime}+2\dot{\xi}\dot{\xi}^{\prime}\Big)=-\gamma\Big(\dot{X} + \dot{\xi} + \dot{\xi}\xi^{\prime} + \xi\dot{\xi}^{\prime}\Big)
\nonumber\\ 
-(1-\xi^{\prime}-\xi^{\prime 2}-\xi\xi^{\prime\prime})\Big\lbrack U^{\prime}(X)
\nonumber \\
+U^{\prime\prime}(X)\xi+U^{\prime}(X)\xi^{\prime}+U^{\prime\prime}(X)\xi\xi^{\prime}+U^{\prime}(X)\xi^{\prime 2}
\nonumber \\
+U^{\prime}(X)\xi\xi^{\prime\prime}+\frac{1}{2}U^{\prime\prime\prime}(X)\xi^{2}+U^{\prime\prime}(X)\xi\xi^{\prime}\Big\rbrack+F(X,t)
\nonumber \\
+ \xi F^{\prime}(X,t) +\xi\xi^{\prime}F^{\prime}(X,t) +\frac{1}{2}\xi^{2}F^{\prime\prime}(X,t).
\end{eqnarray*}
Next, after performing time averaging we ultimately obtain
\begin{eqnarray}
 m\ddot{X}+\gamma\dot{X}& = & -U^{\prime}(X)+U^{\prime}(X)<\xi^{\prime 2}>
\nonumber \\
&-&\frac{1}{2}\frac{\partial}{\partial X}(U^{\prime\prime}(X)<\xi^{2}>)+<\xi F^{\prime}(X,t)>
\nonumber \\
&=& -\frac{\partial U_{eff}}{\partial X},
\end{eqnarray}
with $U_{eff}(X) = U(X)+U^{1}(X)+U^{2}(X)$, where U(X) is the systematic periodic potential, $U^{1}(X)$ is the first order correction to the effective potential given by Eq.(18) and the second order correction to the effective potential is given by
\begin{eqnarray}
U^{2}(X)& = & \frac{1}{4m^2\omega^2(\omega^2+\frac{\gamma^2}{m^2})}\Big\lbrack(f^2(X)+g^2(X))U^{\prime\prime}(X)
\nonumber \\
&-& 8\int^{X}dy\Big(f^{\prime 2}(y)+g^{\prime 2}(y)\Big)\Big\rbrack.
\end{eqnarray}
These results are identical to those obtained by Dutta and Barma \cite{dutta} who have however employed a Fokker-Planck equation approach. In the next section we further extend this analysis by using these results for calculating the effective diffusion enhancement and transport current.\\
{\section{Numerical Scheme and Results}}
We numerically solve the following two dimensionless equations of motion
\begin{eqnarray}
\ddot{x} + b\dot{x} + V^{\prime}(x) & = & a(x)\cos(\omega t) + \sqrt{2bD_0}\eta(t)
 \\
\ddot{x}+ b\dot{x} & = & -U_{eff}^{\prime}(x) + \sqrt{2bD_0}\eta(t),
\end{eqnarray}
with the aid of the Heun scheme which is basically the Runge-Kutta algorithm. Our main interest as emphasized earlier is to calculate the effective diffusion coefficient which is defined as
\begin{equation}
D_{eff} = \lim_{t\rightarrow\infty}\frac{<<x^2(t)>> - <<x(t)>>^2}{2t},
\end{equation}
where the two brackets respectively denote averages over the initial conditions of position and velocity and over all realizations of thermal noise. We calculate this quantity for the orginal dynamics and effective dynamics for the two special cases: (a) space dependent external periodic force and (b) constant amplitude external periodic force. The results are shown in Fig.3.
\begin{figure}[h]
{\rotatebox{270}{\resizebox{6cm}{9cm}{\includegraphics{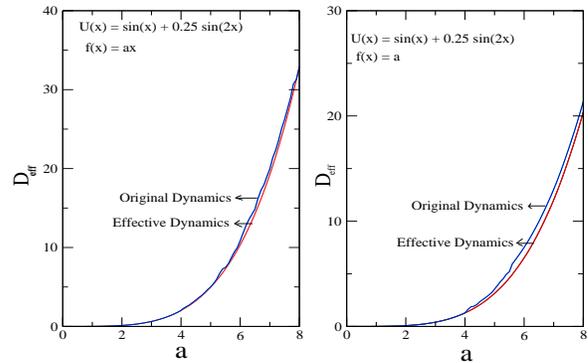}}}}
\caption{The effective diffusion coefficient for original and effective dynamics for two cases (a) space dependent (b) space independent external periodic force.}
\end{figure}
In both the cases the enhancement of effective diffusion coefficient as a function of the amplitude `a' is clearly noticeable. The important difference is that the enhancement is much more pronounced for the space dependent external periodic force than the constant amplitude external periodic force. \\
\noindent
It is known that there are two states of a driven Brownian dynamical system: the locked states, in which the particle stays inside one potential well, and the running states, for which the particle runs over the potential barriers. The first regime is characteristic of a small driving force strength. When the amplitude of the external field is made large, running states appear where we can see both diffusive and regular behavior of the particle. We have shown above that the effective diffusion coefficient increases rapidly when the externally applied force exceeds a certain critical strength. This leads to the existence of an optimal ``a" for the enhancement of diffusion rate.  This phenomenon is reminiscent of Stochastic Resonance (SR) \cite{benzi,nicol,fauv,mcna,gamma,jung1,dyk}. So we can hereby employ the acronym ``SR" to imply  accelaration of diffusion. By this we mean that a new diffusion mechanism, with combined action of noise, spacially periodic potential and time-periodic modulation can be more effective than that of a free Brownian motion, since $D_{eff}$ is shown to exceed unity in a large region around some optimal parameter regions. In these regions the optimal matching of the periodic force and the noise drive the particles up to the potential hills from where they scatter at the potential barriers and finally diffuse very quickly into wide regions. For small `a' it is very difficult to push them up to the potential hills. Therefore, in order to get the above mentioned diffusion enhancement we need the optimal collective actions of three forces: spatially periodic gradients, time-periodic modulation and stochastic stimulation. Further we should emphasize that the extra terms in effective potentials due to space dependence of the external force do indeed aid this diffusion enhancement mechanism.\\
\noindent
We have also studied the current `J' which is defined as the time average of the average velocity over an ensemble of initial conditions. Thus it involves two different averages, the first is over M initial conditions, which we take randomly centered around the origin and with an initial velocity equal to zero. For fixed time $t_j$ we calculate the average velocity $v_j = \frac{1}{M}\sum_{i=1}^{M}\dot{x}_i(t_j)$. The second average is over time and yields $J = \frac{1}{N}\sum_{j=1}^{N}v_j$. All quantities of interest are averaged over 250 different trajectories and $10^4$ periods. There are three dimensionless parameters a, b and $\omega$ defined earlier in terms of physical quantities. We vary the parameter 'a' and fix $b = 0.1$ and $\omega = 0.67$ throughout. In Fig.4 we have shown the behavior of the transport currents in case of space dependent external force. Initially the current is zero, following which it increases and peaks at some optimal values of `a', then decreases with the increase of `a'. The following explanation will help to understand the behavior of current. At very low force strength, escape jumps between the neighbouring wells are very rare i.e. the average directed current is very small.
\begin{figure}[h]
{\rotatebox{270}{\resizebox{4cm}{7cm}{\includegraphics{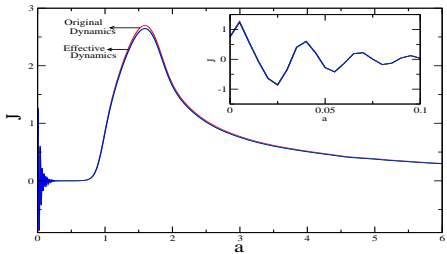}}}}
\caption{The current J of the inertial rocked Brownian motor when external force is space dependent. Inset shows the graph for very low strength force (`a').}
\end{figure}
The input energy is mostly expanded into the kinetic energy of the intrawell motion and eventually dissipates. As `a' is increased further, the Brownian motor mechanism starts to work and some part of energy contributes to the net motion of the particle. Now due to inertia, the mean velocity increases and reaches a maximum. Then `a' reaches a second threshold value above which the current starts to decrease because of the debilitating effect of the ratchet potential. The occurrence of multiple reversals of the directed current, as is shown in the inset of Fig.4, is an interesting feature of the inertial Brownian motor system \cite{jung2,mat,barbi,borro,larro}. The phenomenon of current reversals can be described by different stability properties of the perturbed rotating orbits of the system \cite{barbi}.\\
\begin{figure}[h]
{\rotatebox{270}{\resizebox{4cm}{7cm}{\includegraphics{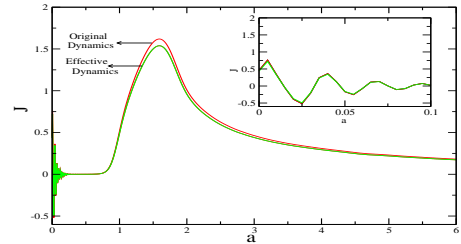}}}}
\caption{The current J of the inertial rocked Brownian motor when external force is space independent. Inset shows the graph for very low strength force (`a').}
\end{figure}
By comparing the figures (4) and (5) we can surmise that the current is much more substantial for the space dependent external force case. Extra terms in the effective potential arising from space dependence of the external force does help in increasing the current.\\ 
{\section{Summary and Conclusions}}
In this section we present an overview of the principal results of this paper.  We have addressed the problem of underdamped Brownian particle in a position dependent periodic driving force in the high frequency regime. We have then calculated the effective potential upto second order in the expansion parameter $\xi$ and used these results to calculate the effective diffusion coefficient and transport current. In the high frequency regime the particle makes small but rapid excursions around a smooth path along which the motion is relatively slow. A systematic perturbative treatment in powers of the excursion amplitude shows that the first order correction in the effective potential exists only if the externally applied rapidly oscillating field is space dependent. This first order correction term (Eq.(18)) is the average kinetic energy which contributes to the work done against damping. The second order correction to the effective potential shows that a nontrivial contribution arises even for position independent driving.\\
\noindent
We have employed our derived results for the calculation of the effective diffusion coefficient and transport current. We obtained the effective diffusion coefficient by solving both the orginal dynamics and effective dynamics. We noted a giant enhancement of diffusion and the results arising from original and effective dynamics agree very well. This validates our method of calculation in the high frequency regime. The enhancement of diffusion is a result of the optimal collective actions of spatially periodic gradients, time periodic modulation and thermal noise. The enhancement is much more pronounced for the space dependent periodic external force which can be understood in terms of the extra terms arising in the effective potential from the space dependence of external force. We  have analyzed the transport properties and the behavior of current for the Brownian motor mechanism and compared the currents for two cases: space dependent and space independent external forces. The current is much larger for the space dependent case.\\
\noindent
Finally we would like to emphasize once again the practical implications of this work. The parameters specifying the particles enter the effective potential, which can be used to separate different species of Brownian particles by identifying the minima of the effective potential. One can control the diffusion rates by varying periodic spatial gradients and the space dependent external field. In addition to myriad applications mentioned earlier, systems described by Eq.(1) are realized for charged particles moving on thermal surfaces under periodic potentials subjected to time varying fields. Recent motivation to study these  systems has been inspired by the theoretical modeling of the molecules called kinesin and myosin, which posess the ability to move unidirectionally along structural filaments of microtutubulin and actin.\\ 
\section*{Acknowledgments}
The authors would like to thank J. Garcia-Palazios for helpful discussions. M.B. also acknowledges financial support from CSIR. M.S. thanks the S. N. Bose Centre for a Summer Student support.

\end{document}